\documentclass{ifacconf}
\usepackage{amsmath}
\usepackage{natbib}
\usepackage{mathtools}
\usepackage{amssymb}

\usepackage{amsfonts}
\usepackage{graphicx}
\usepackage{subfigure}
\usepackage{color}

\renewcommand{\qed}{\hfill\blacksquare}
\newcommand{\qedwhite}{\hfill \ensuremath{\Box}}

\newtheorem{definition}{Definition}
\newtheorem{remark}{Remark}
\newtheorem{exam}{Example}

\newcommand{\R}{\mathbb{R}}
\newcommand{\N}{\mathbb{N}}
\newcommand{\C}{\mathcal{C}}
\newcommand{\X}{\mathcal{X}}
\newcommand{\LL}{\ell}

\newcommand{\D}{\mathcal{D}}
\newcommand{\Sig}{\mathcal{S}}
\newcommand{\OO}{\Omega}
\newcommand{\Ob}{\mathcal{O}}
\newcommand{\sat}{\vDash}

\begin{document}
\begin{frontmatter}

\title{Assume/Guarantee Contracts for Dynamical Systems:\\ Theory and Computational Tools\thanksref{footnoteinfo}} 

\thanks[footnoteinfo]{This work was supported by DENSO Automotive Deutschland GmbH.}

\author[KTH]{Miel~Sharf} 
\author[RUG]{Bart~Besselink} 
\author[DNDE]{Adam~Molin}
\author[DNMI]{Qiming~Zhao}
\author[KTH]{Karl~Henrik~Johansson}

\address[KTH]{Division of Decision and Control Systems, KTH Royal Institute of Technology, and Digital Futures. 10044 Stockholm, Sweden (e-mail: {\rm\{sharf,kallej\}@kth.se)}.}
\address[RUG]{Bernoulli Institute for Mathematics, Computer Science and Artificial Intelligence, University of Groningen, 9700 AK Groningen, The Netherlands (e-mail: {\rm b.besselink@rug.nl}).}
\address[DNDE]{DENSO Automotive Deutschland GmbH, Freisinger Str. 21-23, 85386 Eching, Germany (e-mail: {\rm a.molin@eu.denso.com}).}
\address[DNMI]{DENSO International America, Inc., 24777 Denso Dr., Southfield, MI 48033, USA (e-mail: {\rm qiming.zhao@na.denso.com}).}

\begin{abstract}                
Modern engineering systems include many components of different types and functions. Verifying that these systems satisfy given specifications can be an arduous task, as most formal verification methods are limited to systems of moderate size. Recently, contract theory has been proposed as a modular framework for defining specifications. In this paper, we present a contract theory for discrete-time dynamical control systems relying on assume/guarantee contracts, which prescribe assumptions on the input of the system and guarantees on the output. We then focus on contracts defined by linear constraints, and develop efficient computational tools for verification of satisfaction and refinement based on linear programming. We exemplify these tools in a simulation example, proving a certain safety specification for a two-vehicle autonomous driving setting.
\end{abstract}


\end{frontmatter}

\section{Introduction}
Engineering systems are often comprised of many components having different types and functions, including sensing, control, and actuation. Moreover, systems are subject to many specifications, such as safety and performance. Safety specifications can be captured using the notions of set-invariance (\cite{Blanchini2008}), while performance specifications are usually defined using a bound on the gain of the system, or using passivity, both can be captured using the framework of dissipativity (\cite{vanderSchaft2000}).

However, modern systems such as intelligent transportation systems, complex robotics and smart manufacturing systems have more complex specifications which cannot be captured by the safety and dissipativity frameworks, e.g. behaviour, tracking, and temporal logic specifications. Formal methods in control have been developed to address this issue (\cite{Belta2007,Tabuada2009,Wongpiromsarn2010}). This framework can be used to express temporal logic specifications (\cite{Tabuada2006}). Unfortunately, even if we care only about safety, the sheer size of modern engineering systems implies that formal verification methods are ineffective, as the need to discretize the state-space results in a curse of dimensionality. 
Such verification processes can also be extremely wasteful, as even a minuscule change to the dynamical system (e.g., a small change in one of its components) requires starting the verification processes from scratch.

In this paper we present a verification approach relying on contract theory. Contract theory was first developed in the field of software engineering as a modular approach to system design (\cite{Meyer1992}), and it has proved useful for design of cyber-physical methods, both in theory and in practice (\cite{Nuzzo2014,Nuzzo2015,Naik2020,PhanMinh2019}). Contracts prescribe assumptions on the environments a software component can act in, and guarantees on its behaviour in those environments (\cite{Benveniste2007,Benveniste2018}). The two main approaches for contract theory in computer science include assume/guarantee contracts, which put assumptions on the input to a software component and prescribe guarantees on its output, and interface theories, which provide specifications on the interaction of a component with its environment. In both cases, computational tools for verifying that a given component satisfies a given contract are needed in order to apply the theory.

In recent years, some attempts were made to define a contract theory for dynamical (control) systems. An assume/guarantee framework for continuous-time dynamical systems based on the notion of simulation was considered in \cite{Besselink2019}, in which an algorithm for verifying that a system satisfies a given contract was provided using geometric control theory methods (\cite{vanderSchaft2004}). Assume/guarantee contracts have also been considered in \cite{Saoud2018,Saoud2019}, in which assumptions are made on the input signals and guarantees are on the state and output signals.

In this paper, we present a framework for assume/guarantee contracts prescribing assumptions on the inputs and guarantees on the output, extending the framework of \cite{Saoud2018,Saoud2019}. First, we allow the requirement on the output to depend on the input, which is natural for sensor systems and tasks like tracking. Second, we do not limit the internal structure of the component, for instance, we do not specify its state. This means that the analysis of a composite system can be conducted at a preliminary design stage, before we even know whether, for example, a sensor is a first- or a second-order system, or before we know the controller will be static or not. We also define satisfaction, refinement, and cascaded composition for contracts. We then focus on contracts in which the assumptions and guarantees are prescribed using linear inequalities, and present efficient computational tools for verifying satisfaction and refinement, which are based on linear programming (LP). This is the main contribution of this paper.

The rest of the paper is organized as follows. Section \ref{sec.AG} presents assume/guarantee contracts as well as the notions of satisfaction, refinement, and cascaded composition, and gives examples. Section \ref{sec.Comp} develops computational methods for verification of satisfaction and refinement. Section \ref{sec.Simul} provides a simulation example.
\paragraph*{Notation}
We denote the collection of natural numbers by $\N = \{0,1,2,\ldots\}$. For two sets $X,Y$, we denote their Cartesian product by $X\times Y$.
For a positive integer $n$, we denote the collection of all signals $\N \to \R^n$ by $\Sig^n$. For vectors $v,u \in \mathbb{R}^n$ , we understand $v \le u$ as an entry-wise inequality. Moreover, we denote the Euclidean norm of a vector $v\in \R^n$ as $\|\cdot\|$, and the operator norm of a matrix $P$ as $\|P\| = \sup_{v \neq 0} \frac{\|Pv\|}{\|v\|}$.
Given a state-space system $(A,B,C,D)$, we denote the observability matrix $\mathcal{O}_m = [C^\top,(CA)^\top,\ldots,(CA^m)^\top]^\top$ and define the observability index $\nu$ as the minimal integer such that ${\rm rank}~\mathcal{O}_\nu = {\rm rank}~\mathcal{O}_{\nu+1}$. Moreover, given a state $x$ for the system, we let $p_\Ob(x)$ be the projection of $x$ on the observable subspace of the system.

\section{Assume/Guarantee Contracts}\label{sec.AG}
In this section, we define the class of systems for which we introduce an abstract framework of assume/guarantee contracts, as well as supporting notions such as satisfaction, refinement, and cascaded composition. This is an adaptation of the framework presented in \cite{Benveniste2018}. In section \ref{sec.Comp}, we will specialize to a class of contracts for which efficient computational tools can be introduced.

\begin{definition} \label{def.Systems}
A system $\Sigma$ is a tuple $(\X_0,A,B,C,D)$ with input $d\in \Sig^{n_d}$, output $y \in \Sig^{n_y}$, and state $x\in \Sig^{n_x}$. The set $\X_0 \subseteq \R^{n_x}$ is a set of initial conditions, and $A,B,C,D$ are matrices of appropriate sizes such that the state evolution and output are given by the following equations:
\begin{align} \label{eq.GoverningEquations}
    \begin{cases}
    x(0)&\hspace{-6pt}\in \X_0\\
    x(k+1)&\hspace{-6pt}= Ax(k) + Bd(k),~\forall k \in \N\\
    y(k)&\hspace{-6pt}= Cx(k) + Dd(k),~\forall k \in \N.\\
    \end{cases}
\end{align}
For signals $d\in \Sig^{n_d}$ and $y\in \Sig^{n_y}$, we write $y\in \Sigma(d)$ if there exists a signal $x\in \Sig^{n_x}$ such that $d(\cdot),x(\cdot),y(\cdot)$ satisfy \eqref{eq.GoverningEquations}.
\end{definition}
It is essential to include the set of allowable initial states $\X_0$ in the definition of a system in order to discuss various specifications. For example, asking whether the output of a system lies in a given safe set is meaningless if we make no assumptions on the initial state, no matter the value of the input $d(\cdot)$.
\begin{figure}[b]
    \centering
    \includegraphics[scale = 0.8]{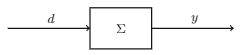}
    \caption{A dynamical system}
    \label{fig.DynSys}
\end{figure}
\begin{remark} \label{rem.InitDepend}
Definition \ref{def.Systems} can be generalized by allowing $\X_0$ to be dependent of $d(0)$. This is reasonable for systems trying to track $d(\cdot)$, assuming their initial tracking error is not too large. This is also reasonable for systems trying to avoid an obstacle whose position is defined by $d(\cdot)$, assuming the system does not start on top of the obstacle.
\end{remark}

We wish to consider specifications on the behaviour of dynamical systems. A dynamical system can be thought of as a map from input signals $d(\cdot) \in \Sig^{n_d}$ to output signals $y(\cdot) \in \Sig^{n_y}$, as in Fig. \ref{fig.DynSys}. As such, we can adopt the formulation of assume/guarantee contracts by merely making assumptions on the input variable $d(\cdot)$ and demanding guarantees on the output variable $y(\cdot)$ given the input $d(\cdot)$.
\begin{definition} \label{defn.AG}
An assume/guarantee contract is a pair $(\D,\OO)$ where $\D \subseteq \Sig^{n_d}$ are the assumptions and $\OO \subseteq \Sig^{n_d} \times \Sig^{n_y}$ are the guarantees.
\end{definition}
In other words, we put assumptions on the input $d(\cdot)$ and demand guarantees on the input-output pair $(d(\cdot),y(\cdot))$. 
\begin{exam}
We say that a system $d\mapsto y$ has finite $\LL_2$-gain no more than $\beta$ if for any $d \in \LL_2$, we have $y\in \LL_2$ and $\|y\|_{\rm \LL_2} \le \beta \|d\|_{\rm \LL_2}$. This property can be written as an assume/guarantee contract $\C = (\D,\OO)$, where $\D = \LL_2$ and $\OO = \{(d(\cdot),y(\cdot)): \|y\|_{\rm \LL_2} \le \beta \|d\|_{\rm \LL_2}\}$.
\end{exam}
\begin{exam}
We say that a SISO system $d\mapsto y$ exponentially tracks constant signals with exponent $\lambda \in (0,1)$ if for any constant input $d$, the output $y$ satisfies $|y(k)-d(k)| \le \lambda |y(k-1)-d(k-1)|$ for all $k$. This property can be written as an assume/guarantee contract $\C = (\D,\OO)$, where $\D = \{d(\cdot): d(k+1) = d(k),~\forall k\}$ and $\OO = \{(d(\cdot),y(\cdot)): |y(k+1)-d(k+1)| \le \lambda |y(k)-d(k)|,~\forall k\}$.
\end{exam}

Let us now define the notion of satisfaction. This notion connects systems and assume/guarantee contracts, by defining when a given system satisfies the specifications defined by a given contract.

\begin{definition}
We say that a system $\Sigma$ satisfies $\C = (\D,\OO)$ (or implements $\C$), and write $\Sigma \sat \C$, if for any $d\in \D$ and any $y\in \Sigma(d)$, $(d,y)\in \OO$.
\end{definition}

\begin{exam}
Consider the following contract $\C=(\D,\OO)$: $d,y\in \Sig^{1}$, 
\begin{align*}
\mathcal{D} &= \{d(\cdot): d(k+1) = d(k) \in [-1,1],~\forall k\}\\
\OO &= \{(d(\cdot),y(\cdot)): |y(k) - d(k)| \le 2^{-k},~\forall k\}.
\end{align*}
First, consider a system $\Sigma_1$ with no state (i.e., $x(\cdot) \in \Sig^0$ and $\X_0 = \emptyset$), $A=B=C=0$ and $D = 1$, i.e. $\Sigma_1$ is a static gain $K=1$. Thus, for any signal $d(\cdot)$ we have $\Sigma_1(d) = \{d\}$, implying that for any $d\in \D$ and $y\in \Sigma_1(d)$ we have $(d,y)\in \Omega$, hence $\Sigma_1 \sat \C$.
Second, consider a system $\Sigma_2$ with a state $x(\cdot)\in \Sig^1$ such that $\X_0 = \{0\}$, $A=B=0.5$, $C=1$ and $D=0$. For any $d(\cdot) \in \D$, there exists some $v\in [-1,1]$ such that $d(k) = v,~\forall k$. If we feed $d(\cdot)$ into $\Sigma$, we get an output $y(k) = x(k) = (1-0.5^k)v$. Thus, $|d(k)-y(k)|\le 2^{-k}$, as $|v|\le 1$, and $\Sigma_2 \sat \C$.
\end{exam}

\subsection{Refinement and Composition}
One of the greatest perks of contract theory is its modularity, as one can refine a contract on a composite system by ``smaller" contracts on subsystems, which can be further refined by even ``smaller" contracts on individual components. The two notions supporting this idea are refinement, defining when one contract is stricter than another, and composition, defining the coupling of multiple contracts. In this subsection, we define the notion of refinement for assume/guarantee contracts, as well as a restricted notion of contract composition for cascade systems. Computational tools for these notions will be provided in the next section.
We start by defining refinement:
\begin{definition} \label{def.refine}
Let $\C_i = (\D_i,\OO_i)$ be contracts for $i=1,2$. We say $\C_1$ \emph{refines} $\C_2$ (and write $\C_1 \preccurlyeq \C_2$) if $\D_1 \supseteq \D_2$ and $\Omega_1 \cap (\D_2 \times \Sig^{n_y}) \subseteq \Omega_2 \cap(\D_2 \times \Sig^{n_y})$, where $n_y$ is the dimension of the output.
\end{definition}
Colloquially, $\C_1 \preccurlyeq \C_2$ if $\C_1$ assumes less than $\C_2$, but guarantees more given the assumptions. 

\begin{exam}
Consider two contracts used for tracking. The first, $\C = (\D,\OO)$ defines asymptotic tracking of certain inputs, namely $d,y\in \Sig^m$, $\D \subseteq \Sig^{m}$, and
\begin{align*}
    \Omega = \{(d(\cdot),y(\cdot)): \lim_{k\to \infty} \|d(k)-y(k)\| = 0\}
\end{align*}
The second, $\C^\prime = (\D,\OO^\prime)$, defines exponential convergence, i.e., we take some $\lambda \in (0,1)$ and define:
\begin{align*}
    \Omega^\prime = \{(d(\cdot),y(\cdot)): \|d(k)-y(k)\| \le \lambda\|d(k-1)-y(k-1)\|\}.
\end{align*}
By definition, we have $\C^\prime \preccurlyeq \C$.
\end{exam}

Refinement provides a partial ordering of assume/guarantee contracts, and it is ``harder" to satisfy refined contracts:
\begin{prop} \label{prop.PropertiesOfRefinement}
Let $\C_i = (\D_i,\OO_i)$ be assume/guarantee contracts for $i=1,2,3$ and $\Sigma$ be a system. Then, the following statements hold:
\begin{itemize}
    \item $\C_1 \preccurlyeq \C_1$.
    \item If $\C_1 \preccurlyeq \C_2$ and $\C_2 \preccurlyeq \C_3$ then $\C_1 \preccurlyeq \C_3$.
    \item If $\C_1 \preccurlyeq \C_2$ and $\Sigma \sat \C_1$, then $\Sigma \sat \C_2$.
\end{itemize}
\end{prop}
\begin{pf}
We prove the claims in order. The first immediately follows from Definition \ref{def.refine}.
For the second claim, $\C_1 \preccurlyeq \C_2$, $\C_2 \preccurlyeq \C_3$ imply that $\D_1 \supseteq \D_2 \supseteq \D_3$, and that:
\begin{align*}
\Omega_1 \cap (\D_2 \times \Sig^{n_y}) \subseteq \Omega_2 \cap(\D_2 \times \Sig^{n_y})
\\
\Omega_2 \cap (\D_3 \times \Sig^{n_y}) \subseteq \Omega_3 \cap(\D_3 \times \Sig^{n_y})
\end{align*}
by intersecting both sides of the first equation with $(\D_3 \times \Sig^{n_y})$, and using $\D_3 \subseteq \D_2$, we conclude that $\Omega_1 \cap (\D_3 \times \Sig^{n_y}) \subseteq \Omega_3 \cap(\D_3 \times \Sig^{n_y})$, hence $\C_1 \preccurlyeq \C_3$. 
Lastly, suppose $\Sigma \sat \C_1$ and $\C_1 \preccurlyeq \C_2$, and we show that $\Sigma \sat \C_2$. Take any $d\in \D_2$ and $y\in \Sigma(d)$. As $\C_1 \preccurlyeq \C_2$, we conclude that $d\in \D_1$, so $\Sigma \sat \C_1$ implies that $(d,y)\in \Omega_1$. Noting that $d \in \D_2$ and $\C_1 \preccurlyeq \C_2$, we conclude that $(d,y) \in \Omega_2$. \qedwhite
\end{pf}
Proposition \ref{prop.PropertiesOfRefinement} is important in contract theory, as it shows two key properties. First, if we have an original contract $\C$ and a refined contract $\C^{\prime}$, any system satisfying $\C^{\prime}$ also satisfies $\C$. Second, if we have an original contract $\C$ and a refined contract $\C^{\prime}$, any refinement of $\C^{\prime}$ is also a refinement of $\C$. These properties allow us to refine a contract on a composite system by multiple contracts on the individual subsystems, which can be further refined by a plethora of contracts on the individual components in the system. If each component satisfies its corresponding contract, then the composite system will satisfy the original contract.

\begin{figure}[t]
    \centering
    \includegraphics[width=0.45\textwidth]{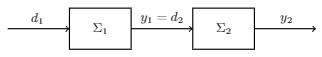}
    \caption{Cascade of contracts}
    \label{fig.Cascade}
\end{figure}

We now move to cascaded composition. Consider the block diagram in Fig. \ref{fig.Cascade}. 
Define the cascaded composition of $\C_1$ and $\C_2$ such that if $\Sigma_1 \sat \C_1$ and $\Sigma_2 \sat \C_2$, the cascade of $\Sigma_1$ and $\Sigma_2$ satisfies the composition $\C_1\otimes\C_2$. First, let us define the cascade of systems:

\begin{definition} \label{defn.CascadeSystem}
Let $\Sigma_i = (\X_i,A_i,B_i,C_i,D_i)$ be systems for $i=1,2$. The cascade $\Sigma_1 \otimes \Sigma_2 = (\X_\otimes,A_\otimes,B_\otimes,C_\otimes,D_\otimes)$ has input $d_\otimes = d_1$, output $y_\otimes = y_2$, state $x_\otimes = [x_1^\top,x_2^\top]^\top$, allowable initial states $\X_\otimes = \X_1 \times \X_2$, and matrices
$A_\otimes = \left[\begin{smallmatrix} A_1 & 0 \\ B_2C_1 & A_2 \end{smallmatrix}\right]$,  $B_\otimes = \left[\begin{smallmatrix} B_1 \\ B_2D_1 \end{smallmatrix}\right]$, $C_\otimes = \left[\begin{smallmatrix} D_2C_1 & C_2 \end{smallmatrix}\right],$ $D_\otimes = D_2D_1$.
\end{definition}

Consider two contracts $\C_i = (\D_i,\OO_i)$ as in Fig. \ref{fig.Cascade}. When defining a contract that is satisfied by the composition, we at least need $d_1\in\D_1$ and $(d_2,y_2)\in\Omega_2$. The latter also requires $d_2\in\D_2$, while the former only implies $(d_1,y_1) \in \OO_1$. This motivates the following definition:
\begin{definition} \label{def.Composition}
For two contracts $\C_i = (\D_i,\OO_i)$, the cascaded composition is $\C_1\otimes\C_2 = (\D_\otimes,\OO_\otimes)$ with input $d_\otimes = d_1$, output $y_\otimes = y_2$,
\begin{align*}
    \D_\otimes &= \{d_\otimes:~ d_\otimes\in \D_1, \left((d_\otimes,y_1)\in \OO_1 \implies y_1 \in \D_2\right)\},\\
    \OO_\otimes &= \{(d_\otimes,y_\otimes): \exists d_2 = y_1, (d_\otimes,y_1)\in \OO_1, (d_2,y_\otimes) \in \OO_2\}.
\end{align*}
\end{definition}

We now prove our main claim about contract composition:
\begin{prop}
Let $\C_1,\C_2$ and $\Sigma_1,\Sigma_2$ be contracts and systems with inputs $d_1,d_2$ and outputs $y_1,y_2$. If $\Sigma_1 \sat \C_1$ and $\Sigma_2 \sat \C_2$, then $\Sigma_1\otimes \Sigma_2 \sat \C_1 \otimes \C_2$.
\end{prop}

\begin{pf}
Denote $\C_i = (\D_i,\OO_i)$.
By Definition \ref{defn.CascadeSystem}, for any $d_\otimes = d_1$ and $y_\otimes = y_2$,  $y_\otimes\in \Sigma_1\otimes\Sigma_2(d_\otimes)$ if and only if there exists some $d_2 = y_1 \in \Sigma_1(d_\otimes)$ such that $y_\otimes \in \Sigma_2(d_2)$.
Now, suppose $d_\otimes = d_1 \in \D_\otimes$ and $y_\otimes = y_2 \in \Sigma_1\otimes\Sigma_2(d_\otimes)$, and let $d_2=y_1$ be such that $d_2 = y_1 \in \Sigma_1(d_\otimes)$ such that $y_\otimes \in \Sigma_2(d_2)$.
As $d_1\in \D_1, y_1 \in \Sigma_1(d_1)$ and $\Sigma_1 \sat \C_1$, we conclude that $(d_1,y_1) \in \OO_1$. As $d_\otimes \in \D_\otimes$, we get $d_2 \in \D_2$. Adding $y_2\in \Sigma_2(d_2)$ and $\Sigma_2 \sat \C_2$ gives that $(d_2,y_2)\in \OO_2$, hence $(d_\otimes,y_\otimes)\in \OO_\otimes$. Thus $\Sigma_1\otimes \Sigma_2 \sat \C_1 \otimes \C_2$. \qedwhite
\end{pf}

\section{Computational Tools for Verification}
\label{sec.Comp}
The previous section presented abstract assume/guarantee contracts for discrete-time dynamical systems, as well the notions of satisfaction, refinement and cascaded composition. In this section, we present computational tools for verifying satisfaction and refinement, relying on mathematical induction and linear programming. We rely on linearity of both the systems and specifications. More precisely, we present computational tools for assumptions of the form $A^1 d(k+1) + A^0 d(k) \le a^0$ for all $k$, and guarantees of the form $G^1 \left[\begin{smallmatrix} d(k+1) \\ y(k+1) \end{smallmatrix}\right] + G^0 \left[\begin{smallmatrix} d(k) \\ y(k) \end{smallmatrix}\right] \le g^0$ for all $k$, where $A^0,A^1,G^0,G^1$ are matrices and $a^0,g^0$ are vectors of appropriate dimensions. Specifications of this form include general bounded signals, as well as outputs of dynamical systems (e.g., the input $d$ is the output of a given first-order system). In Section \ref{sec.Simul}, we use specifications of this form to model a contract where the input is assumed to be a (constrained) trajectory of a dynamical system, and the guarantee is a linear inequality defining safe behaviour.
\subsection{Verifying Satisfaction}
Consider a contract $\C = (\D,\OO)$ where
\begin{align}\label{eq.InductiveD}
    \D &= \{d(\cdot): A^1 d(k+1) + A^0 d(k) \le a^0,~\forall k\},\\ \label{eq.InductiveOmega}
    \OO &= \left\{(d(\cdot),y(\cdot)): G^1 \left[\begin{smallmatrix} d(k+1) \\ y(k+1) \end{smallmatrix}\right] + G^0 \left[\begin{smallmatrix} d(k) \\ y(k) \end{smallmatrix}\right] \le g^0,~ \forall k\right\}
\end{align}
for given $A^0,A^1,G^0,G^1,a^0,g^0$. 
\begin{exam}
Suppose $d(\cdot) \in \Sig^{2}$ is the position of a robot in a field, parameterized by $\mathcal F = \{p\in \R^2: L_1 \le p_1 \le U_1, L_2 \le p_2 \le U_2\}$ for constants $L_1,L_2,U_1,U_2$. Assume that for all $k\in \N$, $d(k) \in \mathcal F$ and $d(k+1) = d(k) + v(k)$ for some velocity $v(k)$ bounded by $V_{\rm max}$, i.e. that $-[V_{\rm max},V_{\rm max}]^\top \le d(k+1) - d(k) \le [V_{\rm max},V_{\rm max}]^\top$. The assumptions are of the form \eqref{eq.InductiveD} where:
\begin{align*}
A^0 = \left[\begin{smallmatrix}1 & -1 & 0 & 0 & 1 & -1 & 0 & 0 \\ 0 & 0 & 1 & -1 & 0 & 0 & 1 & -1 \end{smallmatrix}\right]^\top,~
A^1 = \left[\begin{smallmatrix}0 & 0 & 0 & 0 & -1 & 1 & 0 & 0 \\ 0 & 0 & 0 & 0 & 0 & 0 & 1 & -1 \end{smallmatrix}\right]^\top
\end{align*}
and $a^0 = \left[\begin{smallmatrix} U_1 & L_1 & U_2 & L_2 & -V_{\rm max} & V_{\rm max} & -V_{\rm max} & V_{\rm max} \end{smallmatrix}\right]^\top$.
\end{exam}
Let us make the following assumption on \eqref{eq.InductiveD},\eqref{eq.InductiveOmega}:
\begin{definition}
Given two matrices $V^1,V^0$ and a vector $v^0$, we say $(V^1,V^0,v^0)$ is \emph{extendable} if for any two vectors $u_0,u_1$ and $V^1u_1 + V^0u_0 \le v^0$, there exists some vector $u_2$ such that $V^1 u_2 + V^0 u_1 \le v^0$.
\end{definition}
Assuming $(A^1,A^0,a^0)$ (or $(G^1,G^0,g^0)$) is extendable is not very restrictive. It is equivalent to assuming that any signal $v(\cdot)$ adhering to the assumption, and defined for times $k=0,\ldots,n$, can be extended to a signal defined for all times $k\in \N$ while satisfying the assumption.

\begin{thm} \label{thm.Inductive}
Let $\C = (\D,\OO)$ be a contract with \eqref{eq.InductiveD} and \eqref{eq.InductiveOmega}, and let $\Sigma = (\X_0,A,B,C,D)$ be a system with $x\in \Sig^n$. Assume that $(A^1,A^0,a^0)$ is extendable. Then $\Sigma \sat \C$ if and only if for any $n\in \N$, the following condition holds: for any $d_0,d_1,\ldots,d_{n+1}, x_0,x_1,\ldots,x_{n+1},y_0,y_1,\ldots,y_{n+1}$, the condition:
\begin{align} \label{eq.FullForm}
\begin{cases}
    x_0 \in \X_0,\\
     G^1 \left[\begin{smallmatrix} d_{k+1} \\ y_{k+1} \end{smallmatrix}\right] + G^0 \left[\begin{smallmatrix} d_k \\ y_k \end{smallmatrix}\right] \le g^0,&\forall k=0,\ldots,n-1,\\
    A^1 d_{k+1} + A^0 d_k \le a^0,&\forall k=0,\ldots,n, \\
    x_{k+1} = Ax_k + Bd_k,&\forall k=0,\ldots,n, \\
    y_k = Cx_k + Dd_k,&\forall k=0,\ldots,n+1,
\end{cases}
\end{align}
implies $G^1 \left[\begin{smallmatrix} d_{n+1} \\ y_{n+1} \end{smallmatrix}\right] + G^0 \left[\begin{smallmatrix} d_n \\ y_n \end{smallmatrix}\right] \le g^0$.
\end{thm}

\begin{pf}
Suppose first that whenever \eqref{eq.FullForm} holds we have $G^1 \left[\begin{smallmatrix} d_{n+1} \\ y_{n+1} \end{smallmatrix}\right] + G^0 \left[\begin{smallmatrix} d_n \\ y_n \end{smallmatrix}\right] \le g^0$, and take any $d\in \D$ and $y\in \Sigma(d)$. Note that $A^1d(k+1) + A^0d(k) \le a^0$ holds for all $k$, and for some signal $x(\cdot)$, both $x(k+1) = Ax(k) + Bd(k)$ and $y(k) = Cx(k) + Dd(k)$  hold for all $k$, as well as $x(0) \in \X_0$. Choosing $d_k = d(k), x_k = x(k)$ and $y_k = y(k)$ for all $k = 0,\ldots,n+1$, and using the implication \eqref{eq.FullForm}, we conclude that $G^1 \left[\begin{smallmatrix} d(k+1) \\ y(k+1) \end{smallmatrix}\right] + G^0 \left[\begin{smallmatrix} d(k) \\ y(k) \end{smallmatrix}\right] \le g^0$ can be proved for all $k\in \N$ by induction on the value of $k$, i.e., $\Sigma \sat \C$.

Conversely, suppose $\Sigma \sat \C$, and take $n\in \N$ and some $d_0,x_0,y_0,\ldots,d_{n+1},x_{n+1},y_{n+1}$ such that \eqref{eq.FullForm} holds, and we show that $G^1 \left[\begin{smallmatrix} d_{n+1} \\ y_{n+1} \end{smallmatrix}\right] + G^0 \left[\begin{smallmatrix} d_n \\ y_n \end{smallmatrix}\right] \le g^0$. We show that there exist signals $\hat d(\cdot),\hat y(\cdot)$ such that $\hat{y} \in \Sigma(\hat{d})$, $\hat{d} \in \D$, and $\hat d(k) = d_k, \hat y(k) = y_k$ hold for $k=0,1,\ldots,n+1$. If we show that, we can use $\Sigma \sat \C$ to conclude $(\hat{d},\hat{y}) \in \OO$, implying the desired inequality at time $k=n+1$. Thus, we prove that the signals $\hat d(\cdot),\hat y(\cdot)$ defined above exist.

By extendibility, there exists a signal $\hat{d}\in \D$ such that $\hat d(k) = d_k$ for $k=0,\ldots,n+1$. Choose $\hat{x}(0) = x_0$ and define $\hat x(k+1) = A\hat x(k) + B\hat d(k)$, $\hat y(k) = C\hat x(k) + D\hat d(k)$ for any time $k\in \N$. On one hand, $\hat x(k) = x_k$ and $\hat y(k) = y_k$ hold for all $k=0,\ldots,n+1$. On the other hand, $\hat y \in \Sigma(\hat d)$. We thus showed the existence of $\hat d(\cdot),\hat y(\cdot)$, which implies that the desired implication holds. $\qed$
\end{pf}

The previous theorem allows one to verify that a given system satisfies a given contract by proving a sequence of (infinitely many) implications of the form \eqref{eq.FullForm}. Roughly speaking, this implication guarantees that if the system implements the contract up to time $n$, then it implements the contract up to time $n+1$. Even though this formulation requires infinitely many steps in general, we will soon see that only finitely many implications of the form \eqref{eq.FullForm} needs to be verified. Importantly, the implication \eqref{eq.FullForm} can be cast as an optimization problem. For any $n,p\in \N$ such that $n\ge p$, we consider the following optimization problem:
\begin{align} \label{eq.Prob_np}
    \max ~&~ \max_i~\left[{\rm e}_i^\top \left(G^1 \left[\begin{smallmatrix} d_{n+1} \\ y_{n+1} \end{smallmatrix}\right] + G^0 \left[\begin{smallmatrix} d_n \\ y_n \end{smallmatrix}\right] - g^0\right)\right]\\ \nonumber
    {\rm s.t.} ~& ~G^1 \left[\begin{smallmatrix} d_{k+1} \\ y_{k+1} \end{smallmatrix}\right] + G^0 \left[\begin{smallmatrix} d_k \\ y_k \end{smallmatrix}\right] \le g^0~~~~,\forall k=p,\ldots,n-1,\\ \nonumber
    ~&~A^1 d_{k+1} + A^0 d_k \le a^0~~~~~~~~~~~~,\forall k=p,\ldots,n, \\ \nonumber
    ~&~x_{k+1} = Ax_k + Bd_k~~~~~~~~~~~~~~,\forall k=p,\ldots,n, \\ \nonumber
    ~&~y_k = Cx_k + Dd_k~~~~~~~~~~~~~~~~~,\forall k=p,\ldots,n+1, \\ \nonumber
    ~&~x_p \in \X_p,\\ \nonumber
    ~&~d_k\in \R^{n_d}, x_k\in \R^{n_x}, y_k\in \R^{n_y},\forall k=p,\ldots,n+1,
\end{align}
where $\rm e_i$ are the standard basis elements, and $\X_p$ for $p=1,2,\ldots,n$ are sets to be defined later. We denote this problem as $V_{n,n-p}$ and let $\theta_{n,n-p}$ be its value. Here, $n$ is the last time at which the we know the guarantee holds, $p$ is the first time we consider, and $\ell = n-p$ is the length of history we consider. When taking $p=0$, the problem \eqref{eq.Prob_np} computes the ``worst-case violation" of the guarantee at time $n+1$, given that the guarantees hold up to time $n$. For that reason, Theorem \ref{thm.Inductive} can be restated as:
\begin{cor} \label{cor.Thetann}
Under the assumptions of Theorem \ref{thm.Inductive}, $\Sigma \sat \C$ if and only if $\theta_{n,n} \le 0$ for all $n \in \N$.
\end{cor}

\begin{pf}
$\theta_{n,n} \le 0$ if and only if whenever \eqref{eq.FullForm} holds, $G^1 \left[\begin{smallmatrix} d_{n+1} \\ y_{n+1} \end{smallmatrix}\right] + G^0 \left[\begin{smallmatrix} d_n \\ y_n \end{smallmatrix}\right] - g^0 \le 0$ also holds, which is equivalent to $\Sigma \sat \C$ by Theorem \ref{thm.Inductive}. \qedwhite
\end{pf}

The corollary implies that it suffices to compute $\theta_{n,n}$ for all $n\in\N$ in order to verify $\Sigma\sat \C$. We however prefer to compute $\theta_{n,\ell}$ for small $\ell = n - p$, as this leads to a simpler problem that can be solved more efficiently with existing numerical methods.
The main difficulty in reducing the verification to problems $V_{n,\ell}$ for small $\ell$ is that it requires knowledge of the state trajectory $x(\cdot)$ at time $p = n - l$, captured in \eqref{eq.Prob_np} via the constraint $x_p \in \X_p$. This simply reduces to the initial value $x_0 \in \X_0$ for problems $V_{n,n}$.

An efficient solution of $V_{n,\ell}$ for small $\ell$ requires a characterization of $\X_p$ satisfying the following criteria. First, it is desirable that $\X_p$ is a polyhedral set\footnote{i.e., it is of the form $\{x: Fx \le f\}$ for a matrix $F$ and a vector $f$.}, as \eqref{eq.Prob_np} reduces to a linear problem for which efficient solvers are available, e.g., Yalmip (\cite{Lofberg2004}). 
Second, we would like $\X_p$ to be \emph{independent} of $p$, as this will imply verification of contract satisfaction can be done by solving a finite number of optimization problems (thus not requiring the computation of all $\theta_{n,n}$ as in Corollary \ref{cor.Thetann}). 
Third, $V_{n,\ell}$ is equivalent to $V_{n+1,\ell}$ where $\X_{p+1}$ is the image of $\X_p$ under the dynamics $x_{p+1} = Ax_p + Bd_p$. Combining the last two points, we search for $\X_p$ which is a robust invariant set.

However, these criteria might be contradictory. The last two dictate choosing $\X_p$ as smallest robust invariant set containing $\X_0$, but this set might not be polyhedral even if $\X_0 = \{0\}$ (\cite{Fisher1988}). In fact, the question of whether the minimal robust invariant set containing $\X_0 = \{0\}$ is polyhedral is related to the rationality of the eigenvalues of the matrix $A$ of $\Sigma$. We can try and find some polyhedral robust invariant set containing $\X_0$, not necessarily the smallest one. \cite{Rakovic2005} offer a very partial solution for $\X_0 = \{0\}$, but a general solution is not known to the authors.

To avoid these difficulties, we simply set $\X_p = \R^{n_x}$, at the cost of a more conservative test for contract implementation. Namely, a choice of $\X_p$ that is larger than necessary (i.e., larger than the smallest robust invariant set containing $\X_0$) will make the demand $\theta_{n,\ell}\le 0$ stricter. The following theorem formalizes this case.

\begin{thm} \label{thm.ThetaInducive}
Let $\Sigma=(\X_0,A,B,C,D)$ be a system, and let $\nu$ be its observability index. Take a contract $\C=(\D,\OO)$ such that \eqref{eq.InductiveD} and \eqref{eq.InductiveOmega} hold. Define $\X_p = \mathbb{R}^{n_x}$ for all $p\neq 0$. The following claims hold:
\begin{itemize}
    \item For any $n\in \N, \theta_{n,n} \le \theta_{n,n-1}\le \theta_{n,n-2} \le \ldots \le \theta_{n,0}$. Moreover, for any $\ell \ge 0$, we have $\theta_{\ell,\ell} \le \theta_{\ell+1,
    \ell} = \theta_{\ell+2,\ell} = \theta_{
    \ell+3,\ell} = \cdots$.
    \item Suppose $\D_\star = \{(d_0,d_1): A^1d_1 + A^0d_0 \le a^0\}$ is bounded, and that for any bounded set $E \subseteq \R^{2n_d}$, the intersection of $E\times \R^{2n_y}$ with
    $\Omega_\star =\{(d_0,d_1,y_0,y_1): G^1 \left[\begin{smallmatrix} d_1 \\ y_1 \end{smallmatrix}\right] + G^0 \left[\begin{smallmatrix} d_0 \\ y_0 \end{smallmatrix}\right] \le g^0\}$ is bounded. Then $\theta_{n,\ell} < \infty$ for $n \ge \ell \ge \nu-1$, and $\theta_{n,\ell} = \infty$ if $n,\nu-1 > \ell$.
    \item Given $\ell \ge 0$, if $\theta_{n,n} \le 0$ for any $0 \le n < \ell$ and $\theta_{\ell+1,\ell} \le 0$, then $\Sigma \sat \C$.
\end{itemize}
\end{thm}

\begin{pf}
We prove the claims in order. First, choose any $0\le p \le n-1$. The problem $V_{n,n-p}$ is achieved from $V_{n,n-p+1}$ by changing two constraints. First, we remove the constraints that the guarantees, assumptions and dynamics hold at time $p-1$. Second, while the problem $V_{n,n-p}$ restricts $x_p \in \R^n$, $V_{n,n-p+1}$ restricts $x_p$ to be achieved from the dynamics (via $x_{p-1} \in \R^n$ and $x_p = Ax_{p-1} + Bd_{p-1}$). Thus, $V_{n,n-p+1}$ has the same cost function as $V_{n,n-p}$, but stricter constraints. In particular, as both problems are maximization problems, we conclude $\theta_{n,n-p+1} \le \theta_{n-p}$, as desired. 
Similarly, $V_{n+1,\ell}$ is achieved from $V_{n,\ell}$ by changing the names $d_k,x_k,y_k$ to $d_{k+1},x_{k+1},y_{k+1}$, and changing the set of initial conditions from $\X_0$ to $\R^{n_x}$ (only if $n=\ell$). We can therefore similarly derive $\theta_{n,n} \le \theta_{n+1,n} = \theta_{n+2,n} = \theta_{n+3,n} = \cdots$.

We move to the second claim. By the first part, it suffices to show that $\theta_{n,\nu-1} < \infty$. Consider a feasible solution
$d_{n-\nu+1},x_{n-\nu+1},y_{n-\nu+1},\ldots,d_{n+1},x_{n+1},y_{n+1}$ to $V_{n,\nu-1}$. On one hand, as $\D_\star$ is bounded, $(\D_\star \times \R^{2n_y}) \cap \Omega_\star$ is bounded, meaning that for some constant $M_0>0$, we have $\|d_k\|,\|y_k\|,\|d_{n+1}\| \le M_0$ for $k=n-\nu+1,\ldots,n$. On the other hand, $p_\Ob(x_{n-\nu+1})$ can be achieved as a linear combination of $y_{n-\nu+1},\ldots,y_n$ and $d_{n-\nu+1},\ldots,d_n$ using $\Ob_\nu$. Combining the two, we find a constant $M_1>0$ depending on $M_0$ and $\Ob_\nu$ such that $\|p_\Ob(x_{n-\nu+1})\| \le M_1$. As $\|d_k\| \le M_0$ for all $k$, we get that $\|p_\Ob(x_{n-\nu+k})\| \le M_k$ for $k=1,2,\ldots,\nu+1$, where $M_k = \|A\| M_{k-1} + \|B\|M_0$, hence $\|y_{n+1}\| \le \|C\| M_{\nu+1} + \|D\|M_0$. Thus, the set of feasible $\{d_k,y_k\}_k$ for $V_{n,\nu-1}$ is bounded, and $\theta_{n,\nu-1} < \infty$. One can similarly show that $\theta_{n,\ell} = \infty$ for $\ell \le \nu-2$.

We now consider the third claim. By Corollary \ref{cor.Thetann}, it suffices to show $\theta_{n,n} \le 0$ for all $n \ge \ell$. We prove these inequalities using part 1 of this theorem. First, for $n \ge \ell+1$, we have $\theta_{n,n} \le \theta_{n,\ell} \le \theta_{\ell+1,\ell} \le 0$. Second, for $n = \ell$, we have $\theta_{n,n} = \theta_{\ell,\ell} \le \theta_{\ell+1,\ell} \le 0$. $\qed$
\end{pf}

\begin{remark}
Theorem \ref{thm.ThetaInducive} shows that $\theta_{\ell+1,\ell} = \infty$ if $\ell \le \nu-2$. Thus, we will use the third part of Theorem \ref{thm.ThetaInducive} for $\ell = \nu-1$ to verify implementation.
\end{remark}

\begin{remark}
Consider $V_{n,p}$ for $\X_p = \R^{n_x}$. By using the transfer function associated with the state-space system $(A,B,C,D)$, we can find matrices $E_1,\ldots,E_{m}$ and $F_0,\ldots,F_m$ such that the state-space representation is equivalent to the recursive equation $y(k) = \sum_{r=1}^m E_r y(k-r) + \sum_{r=0}^{m} F_r d(k-r)$.
Thus, $V_{n,\ell}$ for $\ell \ge m$, $p=n-\ell$ and $\X_p = \R^{n_x}$ can be recast as:
\begin{align} \label{eq.Prob_np_nostate}
    \max ~&~ \max_i \left[{\rm e_i}^\top \left(G^1 \left[\begin{smallmatrix} d_{n+1} \\ y_{n+1} \end{smallmatrix}\right] + G^0 \left[\begin{smallmatrix} d_n \\ y_n \end{smallmatrix}\right] - g^0\right)\right]\\ \nonumber
    {\rm s.t.} ~&~ G^1 \left[\begin{smallmatrix} d_{k+1} \\ y_{k+1} \end{smallmatrix}\right] + G^0 \left[\begin{smallmatrix} d_k \\ y_k \end{smallmatrix}\right] \le g^0~,\forall k=p,\ldots,n-1,\\ \nonumber
    ~&~A^1 d_{k+1} + A^0 d_k \le a^0~~~~~~~~,\forall k=p,\ldots,n, \\ \nonumber
    ~&~y_k = \sum_{r=1}^m E_r y_{k-r} + \sum_{r=0}^{m} F_r d_{k-r}\\~&~~~~~~~~~~~~~~~~~~~~~~~~~~~~~~~~~~~~ ,\forall k=p+m,\ldots,n+1, \nonumber \\ \nonumber
    ~&~d_k\in \R^{n_d}, y_k\in \R^{n_y}~~~~~~~~~~,\forall k=p,\ldots,n+1.
\end{align}
This reformulation of \eqref{eq.Prob_np} is more computationally efficient, as it removes a large number of constraints and variables. 
\end{remark}

To conclude this section, we showed one can verify a system $\Sigma$ satisfies a contract $\C$ by solving $\nu + 1$ LP problems, where $\nu$ is the observability index of the system. The first $\nu$ problems assert that $\theta_{n,n} \le 0$ for $n=0,\ldots,\nu-1$, and the last asserts that $\theta_{\nu+1,\nu} \le 0$. The first $\nu$ problems deal with the initial conditions of the system, and the last problem deals with the long-term behaviour of the system. This method can be understood as a version of the k-induction method for model checking (\cite{Donaldson2011}).

\subsection{Verifying Refinement}
In this section, we prescribe computational tools for verifying refinement between contracts defined by linear inequalities. These tools are similar to the ones presented in the work of \cite{Sankaranarayanan2005}.

Consider now two contracts $\C_1 = (\D_1,\OO_1)$ and $\C_2 = (\D_2,\OO_2)$ of the form \eqref{eq.InductiveD} and \eqref{eq.InductiveOmega}, i.e.:
\begin{align} \label{eq.InductiveRefinement}
    \D_1 &= \{d(\cdot): A^1 d(k+1) + A^0 d(k) \le a^0,~\forall k\},\\ \nonumber
    \OO_1 &= \{(d(\cdot),y(\cdot): G^1 \left[\begin{smallmatrix} d(k+1) \\ y(k+1) \end{smallmatrix}\right] + G^0 \left[\begin{smallmatrix} d(k) \\ y(k) \end{smallmatrix}\right] \le g^0,~ \forall k\},\\ \nonumber
    \D_2 &= \{d(\cdot): B^1 d(k+1) + B^0 d(k) \le b^0,~\forall k\},\\ \nonumber
    \OO_2 &= \{(d(\cdot),y(\cdot): H^1 \left[\begin{smallmatrix} d(k+1) \\ y(k+1) \end{smallmatrix}\right] + H^0 \left[\begin{smallmatrix} d(k) \\ y(k) \end{smallmatrix}\right] \le h^0,~ \forall k\},
\end{align}

for some $A^1,A^0,G^1,G^0,B^1,B^0,H^1,H^0,a^0,g^0,b^0,h^0$. We search for a computationally viable way to verify that $\C_1 \preccurlyeq \C_2$. It suffices to show that any $d \in \D_2$ satisfies $d\in \D_1$, and that if $(d,y)\in \OO_1$ and $d\in \D_2$ then $(d,y) \in \OO_2$. As before, we can use inductive reasoning:
\begin{prop} \label{prop.RefinementInductiveImplication}
Let $\C_1,\C_2$ be contracts as in \eqref{eq.InductiveRefinement}, where $G^1 = [G^1_d,G^1_y]$ and $G^0 = [G^0_d,G^0_y]$, and assume both $\left(\left[\begin{smallmatrix} B^1 & 0 \\ G^1_d & G^1_y \end{smallmatrix}\right],\left[\begin{smallmatrix} B^0 & 0 \\ G^0_d & G^0_y \end{smallmatrix}\right], \left[\begin{smallmatrix} b^0 \\ g^0 \end{smallmatrix}\right]\right)$ and $(B^1,B^0,b^0)$ are extendable. $\C_1 \preccurlyeq \C_2$ if and only if the following two implications hold for any $d_0,d_1,y_0,y_1$:
\begin{itemize}
    \item If $B^1 d_1 + B^0 d_0 \le b^0$, then $A^1 d_1 + A^0 d_0 \le a^0$.
    \item If $B^1 d_1 + B^0 d_0 \le b^0$ and $G^1 \left[\begin{smallmatrix} d_1 \\ y_1 \end{smallmatrix}\right] + G^0 \left[\begin{smallmatrix} d_0 \\ y_0 \end{smallmatrix}\right] \le g^0$, then $H^1 \left[\begin{smallmatrix} d_1 \\ y_1 \end{smallmatrix}\right] + H^0 \left[\begin{smallmatrix} d_0 \\ y_0 \end{smallmatrix}\right] \le h^0$.
\end{itemize}
\end{prop}
\begin{pf}
The proof resembles that of Theorem \ref{thm.Inductive}. Suppose first that the two implications hold. We first prove $\D_1 \supseteq \D_2$. Take any $d\in \D_2$, so that $B^1 d(k+1) + B^0 d(k) \le b^0$ holds for all $k$. By assumption, $A^1 d(k+1) + A^0 d(k) \le a^0$ holds for all $k$, and hence $d\in \D_1$. We similarly show that $\OO_1 \cap (\D_2 \times \Sig^{n_y}) \subseteq \OO_2 \cap (\D_2 \times \Sig^{n_y})$.

Conversely, we assume $\C_1 \preccurlyeq \C_2$, and show that both implications hold. Beginning with the first, we take some $d_0,d_1$ such that $B^1 d_1 + B^0 d_0 \le b^0$ holds. By extendibility, there exists a signal $d(\cdot)$ satisfying $d(0)=d_0, d(1) = d_1$ and $B^1 d(k+1) + B^0 d(k) \le b^0$ for all $k$. As $\D_1 \supseteq \D_2$, we have $A^1 d(k+1) + A^0 d(k) \le a^0$ for all $k$. For $k=0$ we get $A^1 d_1 + A^0 d_0 \le a^0$, proving the first implication. The second implication is proved similarly. \qedwhite
\end{pf}

Similarly to the previous subsection, we can verify these implications using linear optimization problems:
\begin{thm} \label{thm.LP_Refinement}
Suppose the assumptions of Proposition \ref{prop.RefinementInductiveImplication} hold. $\C_1 \preccurlyeq \C_2$ if and only if $\psi_\D$ and $\psi_\OO$, the optimal values of the problems below, are non-positive:
\begin{align*}
\psi_\D = \max ~&~ \max_i \left[{\rm e_i}^\top \left(A^1d_1 + A^0d_0 - a^0\right)\right]\\ \nonumber
    {\rm s.t.} ~&~B^1 d_1 + B^0 d_0 \le b^0,~~d_0,d_1\in \R^{n_d}
\end{align*}
\begin{align*}
\psi_\OO = \max ~&~ \max_i \left[{\rm e_i}^\top \left(H^1 \left[\begin{smallmatrix} d_1 \\ y_1 \end{smallmatrix}\right] + H^0 \left[\begin{smallmatrix} d_0 \\ y_0 \end{smallmatrix}\right] - h^0\right)\right]~&~ \\ \nonumber
    {\rm s.t.} ~&~ G^1 \left[\begin{smallmatrix} d_1 \\ y_1 \end{smallmatrix}\right] + G^0 \left[\begin{smallmatrix} d_0 \\ y_0 \end{smallmatrix}\right] \le g^0,~~B^1 d_1 + B^0 d_0 \le b^0\\ \nonumber
    ~&~d_0,d_1\in \R^{n_d},~ y_0,y_1 \in \R^{n_y}
\end{align*}
\end{thm} 
\begin{pf}
Follows from Proposition \ref{prop.RefinementInductiveImplication}, as the implications hold if and only if $\psi_\D$ and $\psi_\OO$ are non-positive. $\qed$
\end{pf}

\if(0)
\subsection{Verifying Complete Compatibility for Cascade}
We defined the cascade composition $\C_1 \otimes \C_2$ of contracts $\C_1 = (\D_1,\OO_1)$ and $\C_2 = (\D_2,\OO_2)$, where $d_2 = y_1$ according to Definition \ref{def.Composition}. As discussed before definition \ref{def.CompletelyCompatible}, it is essential to assure the contracts are completely compatible, i.e. that $\D_\otimes = \D_1$.
In this subsection, we show that if the contracts $\C_1,\C_2$ are of the form \eqref{eq.InductiveRefinement}, one can assert $\C_1,\C_2$ are completely compatible using LP, and in that case, the cascade $\C_1\otimes \C_2$ is also defined using linear inequalities.
We start with the following lemma.
\begin{lem} \label{lem.MappingLinear}
Let $F_A,F_B,E$ be matrices and $f_A,f_B$ be vectors. Define two sets $S_A = \{w \in \R^{m_B}: F_Aw \le f_A\}$ and $S_B = \{w \in \R^{m_A}: F_Bw \le f_B\}$. The image of $S_A$ under the matrix $E$ is contained in $S_B$ if and only if there exists a matrix $P$ such that the following linear constraints hold:
\begin{align*}
    P \ge 0,~~~
    Pf_A \le f_B,~~~
    PF_A = F_B E
\end{align*}
\end{lem}
\begin{pf}
Suppose first that such matrix $P$ exists. Take an arbitrary $x\in S_A$, i.e. $F_A x \le f_A$, and consider $z= Ex$ so that, $F_Bz = F_BEx$. By using $PF_A = F_BE$, we get $F_Bz = PF_Ax$. As $P\ge 0$ and $F_Ax \le f_A$, we get $F_Bz \le Pf_A \le f_B$, meaning that $z=Ex \in S_B$. 

For the converse, suppose the image of $S_A$ under $E$ is contained in $S_B$. For each $i=1,\ldots,m_B$, we let $(F_B)_i$ be the $i$-th row of $F_B$ and consider the following problem:
\begin{align*}
    \mu_i = \max_w ~&~ (F_B)_iEw \\
    {\rm s.t.} ~&~ F_A w \le f_A.
\end{align*}
This is an LP problem, and by assumption $\mu_i \le (f_B)_i$. By strong duality, we conclude that: 
\begin{align*}
    \mu_i = \min_{z^\top} ~&~ z^\top f_A \\
    {\rm s.t.} ~&~ z^\top F_A = (F_B)_i E,~ z^\top \ge 0.
\end{align*}
Take $z_i^\top$ as the optimal solution to the $i$-th dual problem, and let $P$ be the matrix with rows $z_i^\top$. It's clear that $P\ge 0$ and $PF_A = F_B E$, and as $\mu_i \le (f_B)_i$, we have $Pf_A \le f_B$. \qedwhite
\end{pf}

We prove the following characterization of $\D_\otimes$ and $\OO_\otimes$:
\begin{thm} \label{thm.Cascade}
Consider two contracts $\C_1 = (\D_1,\OO_1)$, $\C_2 = (\D_2,\OO_2)$ of the form \eqref{eq.InductiveRefinement}, compatible for cascade. Write $G^i = [G_d^i,G_y^i]$ and $H^i = [H_d^i,H_y^i]$ for $i=0,1$, and assume $\left(\left[\begin{smallmatrix} G_d^1 & G_y^1 \\ A^1 & 0 \end{smallmatrix}\right], \left[\begin{smallmatrix} G_d^0 & G_y^0 \\ A^0 & 0 \end{smallmatrix}\right], \left[\begin{smallmatrix} g^0 \\ a^0\end{smallmatrix}\right]\right)$ is extendable.
\begin{itemize}
    \item $\C_1,\C_2$ are completely compatible if and only if there exist matrices $P_1,P_2$ such that:
    \begin{align*}
& P_1 \ge 0,~~~~P_2 \ge 0,~~~~P_1 g^0 \le b^0,~~~~P_2 a^0 \le b^0, \\
		  & P_1 G_d^0 + P_2 A^0 = 0,~~~~P_1 G_d^1 + P_2 A^1 = 0, \\
          & P_1 G_y^0 = B^0,~~~~~~~~~~~~P_1 G_y^1 = B^1
    \end{align*}
    \item If $\C_1,\C_2$ are completely compatible, then the $\C_1 \otimes \C_2 = (\D_\otimes,\OO_\otimes)$ is defined by linear inequalities:\small
    \begin{align*}
        \mathcal{D}_\otimes &= \{d_\otimes: A^1 d_\otimes(k+1) + A^0 d_\otimes(k) \le a^0,~\forall k\}\\
        \OO_\otimes &= \{(d_\otimes(\cdot),y_\otimes(\cdot): J^1 \left[\begin{smallmatrix} d_\otimes(k+1) \\ y_\otimes(k+1) \end{smallmatrix}\right] + J^0 \left[\begin{smallmatrix} d_\otimes(k) \\ y_\otimes(k) \end{smallmatrix}\right] \le j^0\}, \nonumber
    \end{align*}
\normalsize where we recall that $y_\otimes = [y_1^\top,y_2^\top]^\top$ and:
\begin{align*}
J^1 = \left[\begin{smallmatrix} G_1^d & G^1_y & 0 \\ 0 & H^1_d & H^1_y \end{smallmatrix}\right], ~~
J^0 = \left[\begin{smallmatrix} G_0^d & G^0_y & 0 \\ 0 & H^0_d & H^0_y \end{smallmatrix}\right] ,~~
j^0 = \left[\begin{smallmatrix} g^0 \\ h^0 \end{smallmatrix}\right]
\end{align*}
\end{itemize}
\end{thm}
\begin{pf}
We prove the claims in order. Define:
\begin{align*}
    S_A &= \left\{\left[\begin{smallmatrix} d \\ d^+ \\ y \\ y^+ \end{smallmatrix}\right] : \left[\begin{smallmatrix} G_d^0 & G_d^1 & G_y^0 & G_y^1 \\ A^0 & A^1 & 0 & 0 \end{smallmatrix}\right] \left[\begin{smallmatrix} d \\ d^+ \\ y \\ y^+ \end{smallmatrix}\right] \le \left[\begin{smallmatrix} g^0 \\ a^0 \end{smallmatrix}\right]\right\}, \\
    S_B &= \left\{\left[\begin{smallmatrix}  y \\ y^+ \end{smallmatrix}\right] : \left[\begin{smallmatrix} B^0 & B^1 \end{smallmatrix}\right] \left[\begin{smallmatrix} y \\ y^+ \end{smallmatrix}\right] \le b^0\right\}.
\end{align*}
It's clear $d_2(\cdot) \in \D_2$ if and only if $(d_2(k),d_2(k+1)) \in S_B$ for all $k$, and that both $(d_1(\cdot),y_1(\cdot)) \in \OO_1$  $d_1(\cdot) \in \D_1$  hold if and only if $(d_1(k),d_1(k+1),y_1(k),y_1(k+1)) \in S_A$ for all $k$.  Moreover, $\C_1$ and $\C_2$ are completely compatible if and only if $(d_1,y_1)\in\OO_1$ and $d_1 \in \D_1$ imply that $d_2=y_1 \in \D_2$. Due to extendibility, it's equivalent to show that for any $[d^\top,(d^+)^\top,y^\top,(y^+)^\top]^\top\in S_A$, we have $[y^\top,(y^+)^\top]^\top\in S_B$, i.e. the image of $S_A$ under  $E = \left[\begin{smallmatrix} 0 & 0 & I & 0 \\ 0 & 0 & 0 & I \end{smallmatrix}\right]$ is inside $S_B$. By Lemma \ref{lem.MappingLinear}, it is equivalent to finding a matrix $P$ satisfying:
\begin{align*}
    P \ge 0,~
    P\left[\begin{smallmatrix} g^0 \\ a^0 \end{smallmatrix}\right] \le b^0,~
    P\left[\begin{smallmatrix} G_d^0 & G_d^1 & G_y^0 & G_y^1 \\ A^0 & A^1 & 0 & 0 \end{smallmatrix}\right]  = \left[\begin{smallmatrix} B^0 & B^1 \end{smallmatrix}\right]  \left[\begin{smallmatrix} 0 & 0 & I & 0 \\ 0 & 0 & 0 & I \end{smallmatrix}\right]
\end{align*}
We deduce the desired conditions by writing $P = [P_1,P_2]$. As for the second part of the theorem, if $\C_1,\C_2$ are completely compatible then $\D_\otimes = \D_1$. Thus, the description of $\D_\otimes$ is clear, and the description $\OO_\otimes$ follows from its definition, as $(d_\otimes,y_\otimes) \in \OO_\otimes$, where $d_\otimes = d_1$ and  $y_\otimes = [y_1^\top,y_2^\top]^\top$, if and only if $(d_1,y_1) \in \OO_1$ and $(y_1,y_2)\in \OO_2$. $\qed$
\end{pf}
\fi
To conclude this section, we showed that for contracts defined by time-independent linear inequalities, satisfaction and refinement can be verified using linear programming.
\section{Simulation Example}
\label{sec.Simul}
We exemplify the computational tools prescribed in Section \ref{sec.Comp} using case studies.

\subsection{Contract Satisfaction}
\begin{figure}[b]
    \centering
    \vspace{5pt}
    \includegraphics[width = 0.45\textwidth]{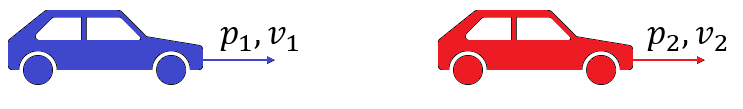}
    \caption{The two-vehicle scenario of Section \ref{sec.Simul}}
    \label{fig.TwoCar}
\end{figure}

Consider two vehicles driving along a single-lane highway, as in Fig. \ref{fig.TwoCar}. We are given a headway $h>0$, and our goal is to verify that the follower keeps at least the given headway from the leader. Denoting the position and velocity of the follower as $p_1(k)$, $v_1(k)$, and the position and velocity of the leader as $p_2(k),v_2(k)$, we want to show that $p_2(k) - p_1(k) - hv_1(k) \ge 0$ holds at any time $k\in \N$. We address this problem using assume/guarantee contracts. 

The input signal to the follower $d(\cdot)$ is $d(k) = [p_2(k),v_2(k)]$. It is reasonable to assume the leader vehicle follows the kinematic laws, i.e.,
\begin{align*}
&p_2(k+1) = p_2(k) + \Delta t v_2(k),\\ & v_2(k+1) = v_2(k) + \Delta ta_2(k),\\
&a_2(k) \in [-a_{\rm min},a_{\rm max}]
\end{align*}
where $a_2(k)$ is the acceleration to the leading vehicle and $\Delta t > 0$ is the length of a discrete time step. As for guarantees, we want to assure that  $p_2(k) - p_1(k) - hv_1(k) \ge 0$ holds for any $k\in \N$. It is clear that these assumptions and guarantees are given by linear inequalities, meaning that the methods of Section \ref{sec.Comp} can be applied. Explicitly, the set of assumptions is of the form \eqref{eq.InductiveD} and the set of guarantees is of the form \eqref{eq.InductiveOmega}, for:
\begin{align*}
&A^1 = \left[\begin{smallmatrix} 1 & 0 \\ -1 & 0 \\ 0 & 1 \\ 0 & -1 \end{smallmatrix}\right],
&&A^0 = \left[\begin{smallmatrix} -1 & -\Delta t \\ 1 & \Delta t \\ 0 & -1 \\ 0 & 1  \end{smallmatrix}\right]
&&a^0 = \left[\begin{smallmatrix} 0 \\ 0 \\ \Delta t a_{\rm max} \\ \Delta t a_{\rm min}\end{smallmatrix}\right], \\
&G^1 = \left[\begin{smallmatrix} 0 & 0 & 0 & 0\end{smallmatrix}\right], 
&&G^0 = \left[\begin{smallmatrix} -1 & 0 & 1 & h\end{smallmatrix}\right], 
&&g^0 = [0].
\end{align*}

We must also specify the system. We assume the follower vehicle also satisfies the kinematic laws, with an acceleration dictated by an affine control law:
\begin{align*}
&p_1(k+1) = p_1(k) + \Delta t v_1(k),~ v_1(k+1) = v_1(k) + \Delta ta_1(k),\\
&a_1(k) = \frac{p_2(k)-p_1(k)}{h\Delta t} - \left(\frac{1}{h} + \frac{1}{\Delta t}\right)v_1(k)  + \frac{v_2(k)}{h} - 1_{\rm m/s^2},
\end{align*}
In other words, the follower can be modeled by a system $\Sigma$
defined by the equations $x(k+1) = Ax(k) + Bd(k) + w$, $y(k) = Cx(k)+Dd(k)$, where $x = y = [p_1,v_1]^\top$, $d = [p_2,v_2]^\top$, $\X_0$ depends on $d(0)$ as we assume the initial state satisfies $p_2(0) - p_1(0) - hv_1(0) \ge 0$ (see Remark \ref{rem.InitDepend}), and the dynamics are given by the matrices:
\begin{align*}
A = \left[\begin{smallmatrix} 1 & \Delta t \\ -\frac{1}{h} & -\frac{\Delta t}{h} \end{smallmatrix}\right],~
B = \left[\begin{smallmatrix} 0 & 0 \\ \frac{1}{h} & \frac{\Delta t}{h} \end{smallmatrix}\right],~
C = I,~D=0,~w=\left[\begin{smallmatrix} 0 \\ -\Delta t \end{smallmatrix}\right]
\end{align*}

\begin{figure}[t]
    \centering
    \subfigure[Velocity of leader] {\scalebox{.33}{\includegraphics{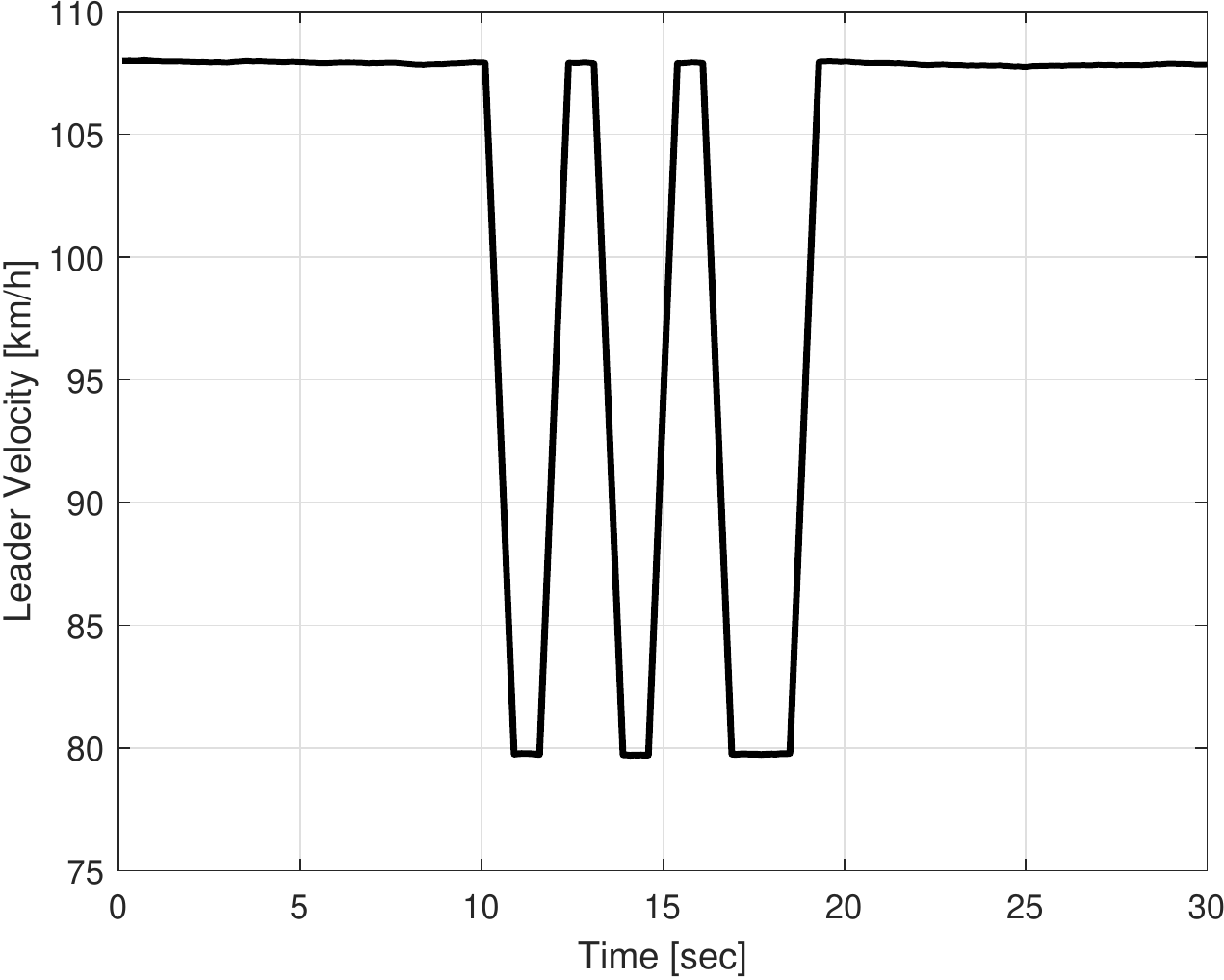}}} \hfill
    \subfigure[Acceleration of leader] {\scalebox{.33}{\includegraphics{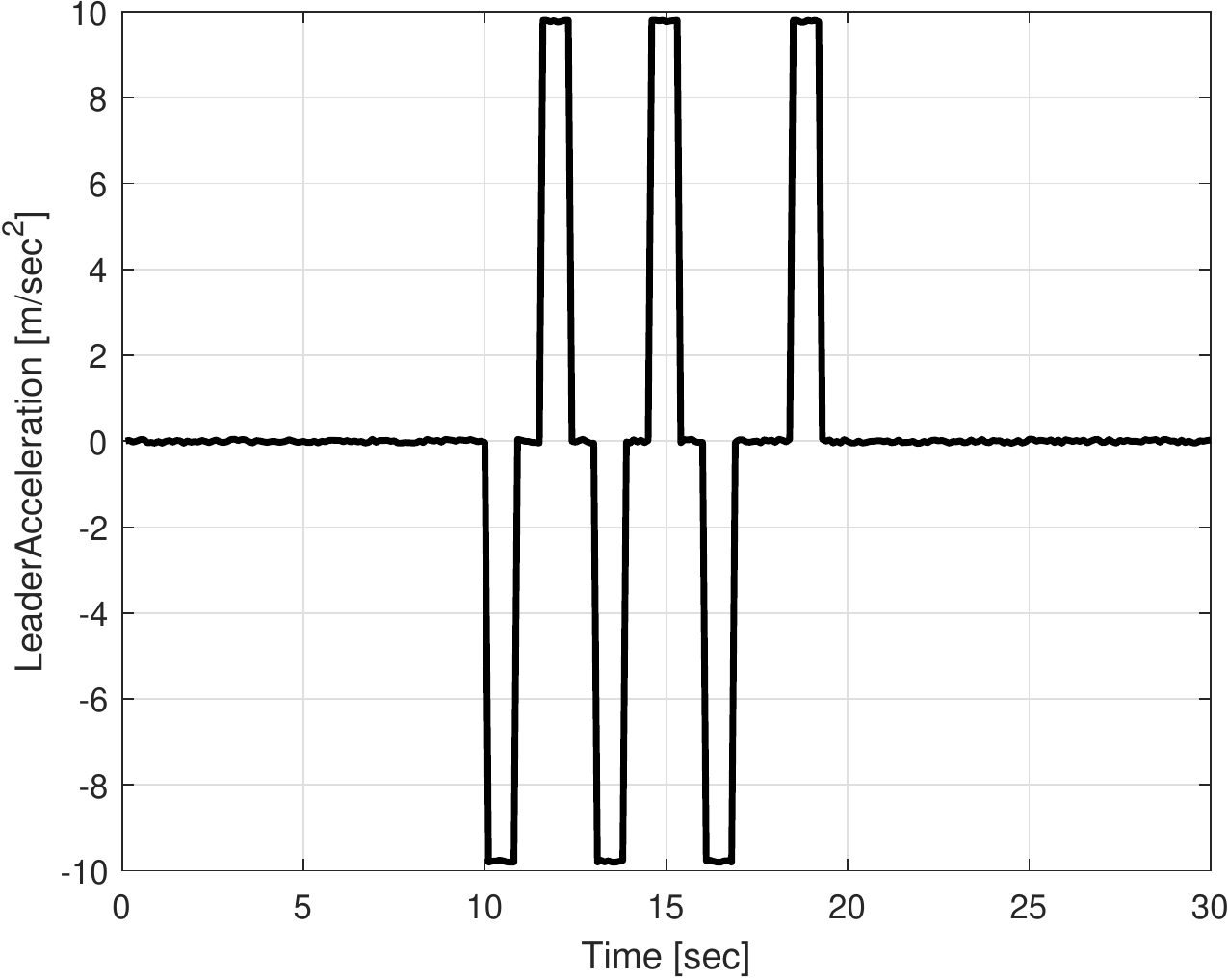}}}
    \caption{Leader vehicle in simulation.}
    \label{fig.Leading}
\end{figure}
\begin{figure}[b]
    \centering
    \subfigure[Velocity of follower vehicle] {\scalebox{.32}{\includegraphics{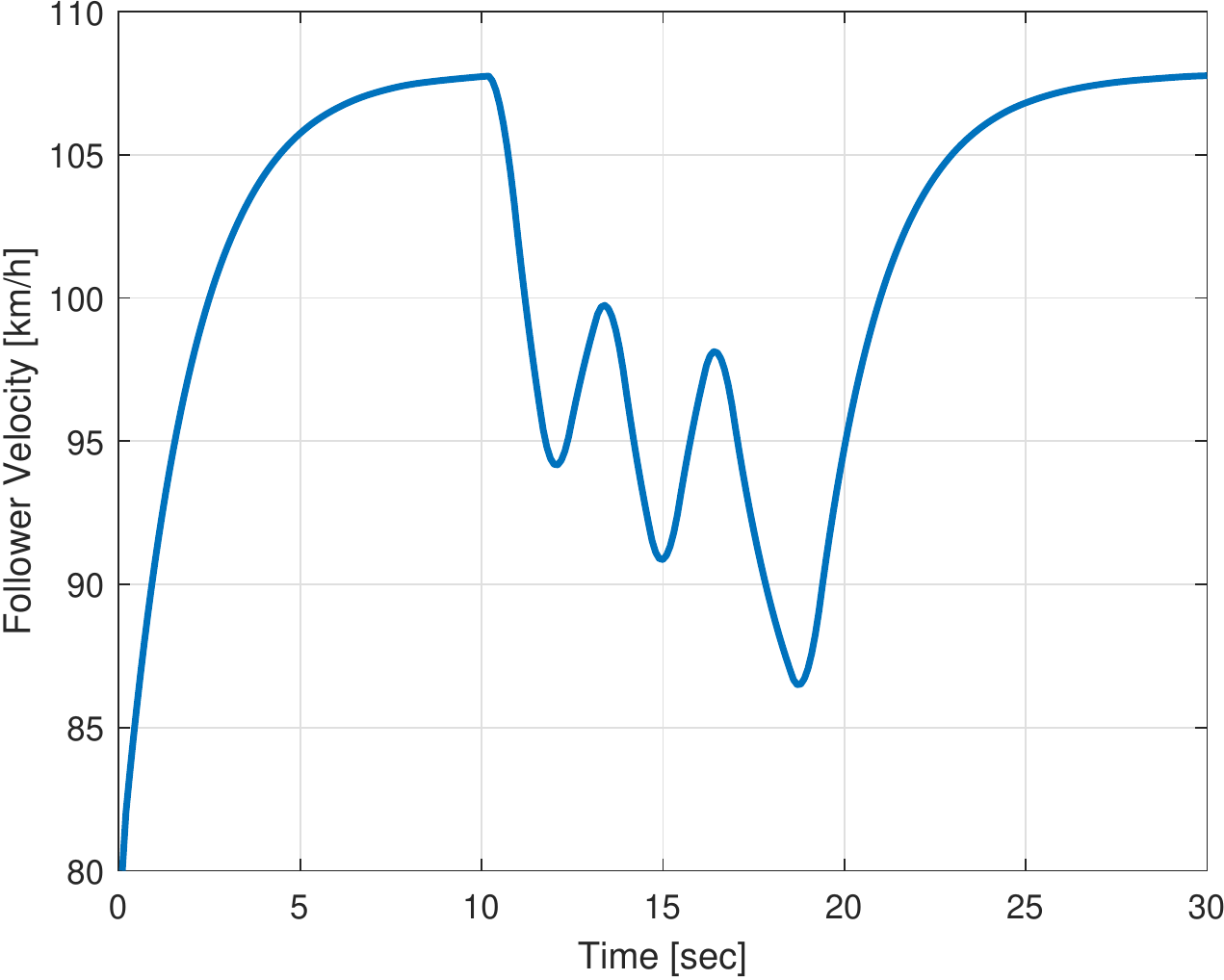}}} \hfill
    \subfigure[Headway between vehicles] {\scalebox{.32}{\includegraphics{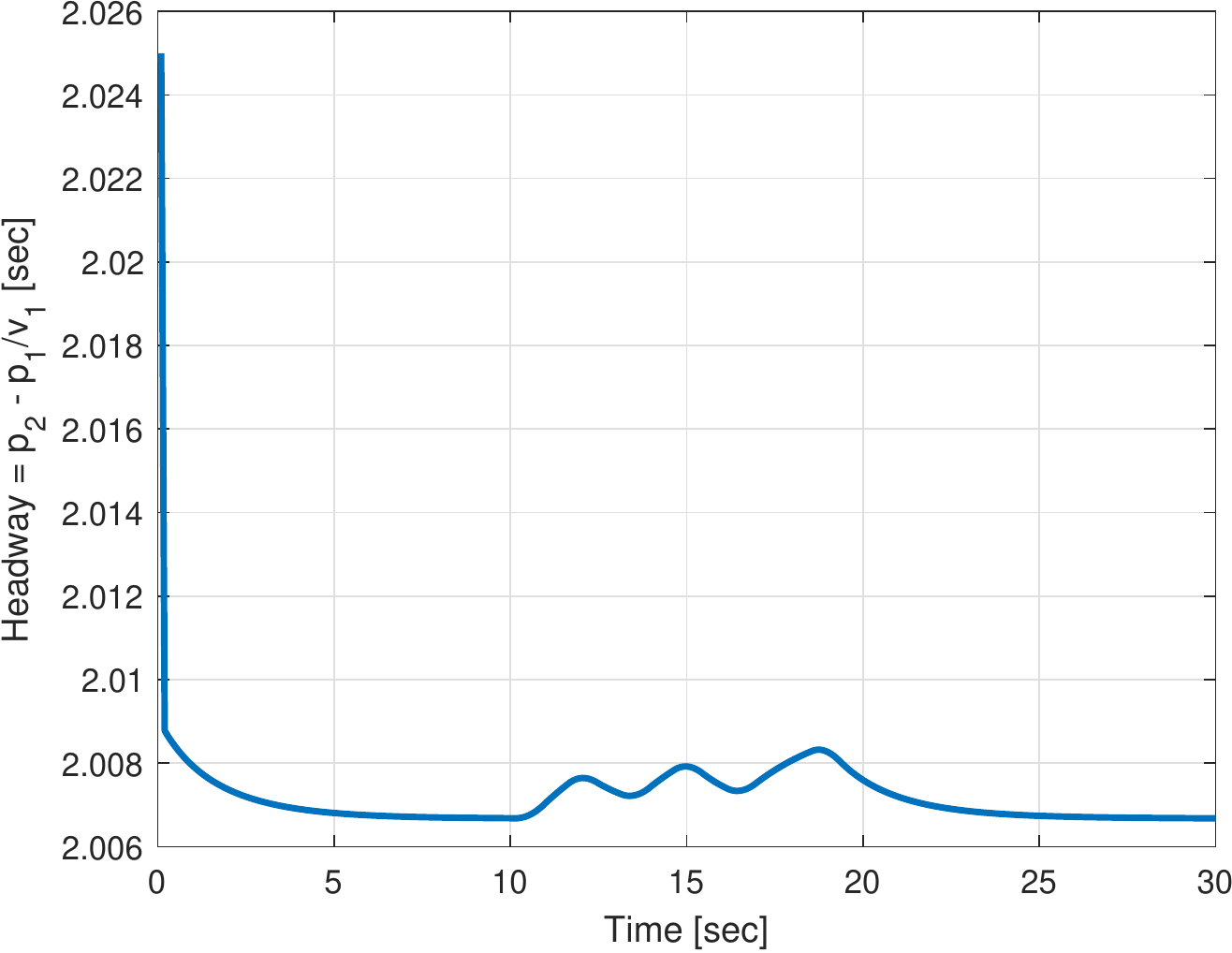}}}
    \caption{Velocity of follower vehicle and headway.}
    \label{fig.Follower}
\end{figure}

We want to prove that $\Sigma \sat \C$ , and we do so using Theorem \ref{thm.ThetaInducive}. The system $\Sigma$ is observable, and its observability index is $\nu = 1$. Thus, it suffices to prove $\theta_{0,0}, \theta_{2,1} \le 0$, where:
\begin{align*}
   \theta_{0,0} =  \max ~&~  -(p_2(0) - p_1(0) - hv_1(0))\\ \nonumber
    {\rm s.t.} ~&~ p_2(0) - p_1(0) - hv_1(0) \ge 0\\ \nonumber
    				~&~ p_1(0),p_2(0),v_1(0),v_2(0) \in \R\\
  \theta_{2,1} =  \max ~&~  -(p_2^+ - p_1^+- hv_1^+)\\ \nonumber
    {\rm s.t.} ~&~  p_2- p_1 - hv_1 \ge 0\\ \nonumber
    ~&~p_2^+ = p_2 + \Delta t v_2,~~ v_2^+ = v_2 + \Delta t a_2 \\ \nonumber
    ~&~ a_2 \in [-a_{\rm min},a_{\rm max}]\\\nonumber
    ~&~p_1^+ = p_1 + \Delta t v_1,~~v_1^+ = v_1 + \Delta t a_1 \\ \nonumber
    ~&~ a_1 = \frac{p_2-p_1}{h\Delta t} - \left(\frac{1}{h} + \frac{1}{\Delta t}\right)v_1  + \frac{v_2}{h}-1\\\nonumber
    ~&~p_2^+,p_1^+,v_2^+,v_1^+,a_2^+,a_1^+ ,p_2,p_1,v_2,v_1,a_2,a_1 \in \R.
\end{align*}
In the problem defining $\theta_{2,1}$, the parameters with ``$+$" correspond to time $k=2$, and the ones without ``$+$" correspond to time $k=1$. We choose parameters $a_{\rm min} = a_{\rm max} = 9.8 {\rm m/s^2}$, $\Delta t = 0.1 {\rm sec}$, $h=2 {\rm sec}$, and solve both LP problems using Yalmip (\cite{Lofberg2004}), computing $\theta_{0,0} = 0, \theta_{2,1} = -0.2$, meaning that $\Sigma \sat \C$ as $\theta_{0,0},\theta_{2,1} \le 0$.

We exemplify that $\Sigma \sat \C$ through simulation. We consider the following trajectory of the leader - its initial speed is about $110{\rm km/h}$, which is roughly kept for 10 seconds. It then starts to sway wildly for 10 seconds between $80{\rm km/h}$ and $110{\rm km/h}$, braking and accelerating as hard as possible. Finally, it stops swaying and keeps its velocity for 10 more seconds. The velocity and acceleration of the leader can be seen in Fig. \ref{fig.Leading}. 
The follower starts $45 \rm m$ behind the leader, so the headway is kept at time $0$. We run the simulation for both vehicles, and plot the headway $\frac{p_2(k)-p_1(k)}{v_1(k)}$  and the velocity of the follower in Fig. \ref{fig.Follower}. It can be seen that the headway is kept throughout the run, so the guarantees are satisfied, as predicted by our analysis. 

\subsection{Contract Refinement}
As in the previous case study, consider the two-vehicle scenario described in Fig. \ref{fig.TwoCar}. As before, we consider contracts about the behaviour of the follower vehicle, where the input $d = [p_2,v_2]$ consists of the position and velocity of the leader vehicle, and the output $y = [p_1,v_1]$ consists of the position and velocity of the follower vehicle. 

We now prescribe two contracts $\C_1,\C_2$ on the follower, where $\C_1 = (\D_1,\OO_1)$ and $\C_2 = (\D_2,\OO_2)$. In both, we assume that the leader vehicle satisfies the kinematic relations, with varying bounds on its acceleration, and guarantee that headway is kept. Namely, for $j=1,2$, the set of assumptions $\D_i$ is given by the following kinematic relations, which must hold for all times $k\in \N$:
\begin{align*}
&p_2(k+1) = p_2(k) + \Delta t v_2(k),\\ & v_2(k+1) = v_2(k) + \Delta ta_2(k),\\
&a_2(k) \in [-a_{{\rm min},j},a_{{\rm max},j}],
\end{align*}
where the parameters $a_{{\rm min},j},a_{{\rm max},j}$ determine the assumed maximum acceleration and deceleration. We also assume that the vehicle is moving forward in both cases, i.e. that $v_2(k) \ge 0$. Similarly, the set $\Omega_j$ is defined by the following headway-preservation safety guarantee:
\begin{align*}
    p_2(k) - p_1(k) - h_j v_1(k) \ge 0,
\end{align*}
where the parameters $h_1,h_2$ determine the desired headway between the vehicles. 

It is clear that if $a_{{\rm min},2} \le a_{{\rm min},1}$ and $a_{{\rm max},2},a_{{\rm max},1}$, then the contract $\C_1$ assumes less than the contract $\C_2$, as its assumptions allow the leading vehicle the accelerate and decelerate more sharply. Moreover, if $h_1 \ge h_2$, then $\C_1$ guarantees more than $\C_2$, as the associated headway is larger. Thus, for this parameter setting, we have that $\C_1 \preccurlyeq \C_2$. We wish to verify this refinement using the tools of Section \ref{sec.Comp}.

First, we note that these contracts are defined by linear inequalities. Specifically, \eqref{eq.InductiveRefinement} holds where the matrices $A^0,A^1,G^0,G^1,B^0,B^1,H^0$ and $H^1$ are given by:
\begin{align*}
&A^1 = \left[\begin{smallmatrix} 1 & 0 \\ -1 & 0 \\ 0 & 1 \\ 0 & -1 \\ 0 & 0 \end{smallmatrix}\right],
&&A^0 = \left[\begin{smallmatrix} -1 & -\Delta t \\ 1 & \Delta t \\ 0 & -1 \\ 0 & 1 \\ 0 & 1 \end{smallmatrix}\right]
&&a^0 = \left[\begin{smallmatrix} 0 \\ 0 \\ \Delta t a_{\rm max,1} \\ \Delta t a_{\rm min,1} \\ 0\end{smallmatrix}\right], \\
&G^1 = \left[\begin{smallmatrix} 0 & 0 & 0 & 0\end{smallmatrix}\right], 
&&G^0 = \left[\begin{smallmatrix} -1 & 0 & 1 & h_1\end{smallmatrix}\right], 
&&g^0 = [0],\\
&B^1 = \left[\begin{smallmatrix} 1 & 0 \\ -1 & 0 \\ 0 & 1 \\ 0 & -1 \\ 0 & 0 \end{smallmatrix}\right],
&&B^0 = \left[\begin{smallmatrix} -1 & -\Delta t \\ 1 & \Delta t \\ 0 & -1 \\ 0 & 1 \\ 0 & 1  \end{smallmatrix}\right]
&&b^0 = \left[\begin{smallmatrix} 0 \\ 0 \\ \Delta t a_{\rm max,2} \\ \Delta t a_{\rm min,2} \\ 0\end{smallmatrix}\right], \\
&H^1 = \left[\begin{smallmatrix} 0 & 0 & 0 & 0\end{smallmatrix}\right], 
&&H^0 = \left[\begin{smallmatrix} -1 & 0 & 1 & h_2\end{smallmatrix}\right], 
&&h^0 = [0].
\end{align*}

By Theorem \ref{thm.LP_Refinement}, $\C_1 \preccurlyeq \C_2$ if and only if the following two optimization problems have a non-positive value:

\begin{align*}
\psi_\D = \max ~&~ c_\D(p_2,p_2^+,v_2,v_2^+)\\ \nonumber
    {\rm s.t.} ~&~ p_2^+ = p_2 + \Delta t v_2, ~ v_2 \ge 0,\\
               ~&~ v_2 - \Delta t a_{{\rm min},2} \le v_2^+ \le v_2 + \Delta t a_{{\rm max},2},\\
               ~&~ p_2,p_2^+,v_2,v_2^+ \in \R.
\end{align*}
\begin{align*}
\psi_\OO = \max ~&~ p_2 + h_2v_2 - p_1\\ \nonumber
    {\rm s.t.} ~&~ p_1 - p_2 - h_1v_2 \ge 0,\\
               ~&~ p_2^+ = p_2 + \Delta t v_2,~v_2 \ge 0\\
               ~&~ v_2 - \Delta t a_{{\rm min},2} \le v_2^+ \le v_2 + \Delta t a_{{\rm max},2},\\
               ~&~ p_1,v_1,p_1^+,v_1^+,p_2,p_2^+,v_2,v_2^+ \in \R,
\end{align*}
where the cost function $c_\D$ is defined as:
\begin{align*}
    c_\D(p_2,p_2^+,v_2,v_2^+) = \max\{&p_2^+ - p_2 - \Delta t v_2,\\
    &p_2+\Delta t v_2 - p_2^+,\\
    &v_2^+ - v_2 - \Delta t a_{{\rm max},1},\\
    &v_2 - \Delta t a_{{\rm min},1} - v_2^+\}
\end{align*}

We choose the parameter $\Delta t = 0.1 {\rm sec}$ for both contracts, the parameters $a_{\rm min,1} = a_{\rm max,1} = 9.8 {\rm m/s^2}$, $h_1=2 {\rm sec}$ for the first contract, and the parameters $a_{\rm min,2} = a_{\rm max,2} = 9.5 {\rm m/s^2}$, $h_2=1.9 {\rm sec}$ for the second contract. We solve both problems using Yalmip (\cite{Lofberg2004}), computing $\psi_\D = -0.03, \psi_\OO = 0$. Because $\psi_\D, \psi_\OO \le 0$, we conclude that $\C_1 \preccurlyeq \C_2$, as expected.

\section{Conclusions and Future Research}
We presented an assume/guarantee contract framework for discrete-time dynamical systems. The framework puts assumptions on the input signal to the system, and prescribes guarantees on the output relative the the input. In particular, as the guarantees do not include the state, systems of different orders can satisfy the same contract. We also defined corresponding fundamental notions such as satisfaction, refinement, and cascaded composition. Perhaps more importantly, we showed that for contracts defined using linear inequalities, satisfaction and refinement can be verified using linear programming, which can be solved efficiently using off-the-shelf optimization software. Finally, we exemplified our methods using a case study on a 2-vehicle leader-follower scenario, where the goal was to obey a certain headway. Future research can extend our results by extending the methods presented in this work for nonlinear, uncertain, or hybrid systems, as well as for verifying compositional refinement, i.e. that a composition of multiple contracts on individual components or subsystems refines a contract on the composite system.

\bibliography{main}   
\end{document}